\documentclass{elsart}

\usepackage{psfig}
\usepackage{epsf}
\usepackage{natbib}

\newcommand{\farcs}{\hbox{$.\!\!{''}$}}
\newcommand{\ltwid}{\mathrel{\raise.3ex\hbox{$<$\kern-.75em\lower1ex\hbox{$\sim$}}}}

\begin{document}

\runauthor{Moran}


\begin{frontmatter}

\title{Radio Properties of NLS1s}

\author{Edward C.\ Moran}
\address{Department of Astronomy, University of California,
Berkeley, CA 94720 USA}

\begin{abstract}
While NLS1s have been studied intensively at X-ray and optical wavelengths,
comparatively little is known about their characteristics in the radio band.
Therefore, we have carried out an investigation of the radio luminosities,
source sizes, spectral index distribution, and variability of a large,
uniformly selected sample of NLS1s.  Our results indicate that, in some
respects, the radio properties of NLS1s differ significantly from those of
classical Seyfert galaxies.  Radio observations of NLS1s may thus provide
important clues regarding the nature of their nuclear activity.
\end{abstract}

\end{frontmatter}


\section{Introduction}
It is well established that the optical and X-ray characteristics of
narrow-line Seyfert~1 galaxies distinguish them from all other types of
active galactic nuclei (AGNs).  Unfortunately, the radio properties of
NLS1s have been less well explored.  In the only study dedicated to the
subject, Ulvestad, Antonucci, \& Goodrich (1995; hereafter UAG) found that
NLS1s do not differ noticeably from nearby classical type~1 and type~2
Seyfert galaxies at centimeter wavelengths, in contrast to the results
obtained in the optical and X-ray bands.  This conclusion was based on
the modest radio powers ($10^{20}$--$10^{23}$ W~Hz$^{-1}$) and small
radio source sizes ($\ltwid 300$~pc) of the objects they examined.  But
as UAG candidly noted, their sample of NLS1s was not defined using a
uniform set of criteria, and only a fraction of the galaxies in it (9/15)
was detected.  We have investigated the radio emission of a larger,
uniformly selected sample of NLS1s in order to gain further insight into
the radio nature of these objects and their relation to other classes of
AGNs.

Our sample of 24 NLS1s is drawn from the catalog of {\sl IRAS\/} sources
detected in the {\sl ROSAT\/} All-Sky Survey (Boller et al.\ 1992; Moran
et al.\ 1996).  Full details regarding the sample definition
are provided in Moran et al.\ (2000).  We have obtained simultaneous
high-resolution A-array VLA observations at 20~cm and 3.6~cm of most of the
{\sl IRAS-} and {\sl ROSAT-}Observed NLS1 (``IRON'') galaxies.  In addition,
nearly all of the objects have been imaged at 20~cm in the moderate-resolution
B and C arrays by Condon et al.\ (1998a) and in the low-resolution D array
as part of the NRAO VLA Sky Survey (Condon et al.\ 1998b).  All but
one of the IRON galaxies are detected at 20~cm; 22 have three or more flux
density measurements at that wavelength.

\section{Population Statistics}
{\it Radio Power Distribution.}  As Figure~1$a$ illustrates, the majority
of the IRON objects have radio powers in excess of $10^{23}$ W~Hz$^{-1}$,
and seven are more luminous than $10^{24}$ W~Hz$^{-1}$---in stark contrast
to the 20~cm radio power distribution for nearby classical Seyfert galaxies
(Ulvestad \& Wilson 1989).  Thus, it would appear that NLS1s are frequently
more luminous than nearby Seyferts in the radio band.  We have also determined
the 20~cm luminosity distribution for 77 classical
Seyfert galaxies in the {\sl IRAS-ROSAT\/} catalog from which the IRON
sample was drawn, based on VLA observations by Condon et al.\ (1998a).  As
indicated in Figure~1$b$, the radio luminosities of the IR/X-ray--selected
Seyferts tend to be higher than those of the nearest classical Seyfert
galaxies, but they do not extend to the very high luminosities displayed
by NLS1s selected the same way.  Interestingly, the radio--to--infrared
and radio--to--X-ray flux ratio distributions of the IRON galaxies and the
{\sl IRAS-ROSAT\/} Seyferts do {\it not\/} differ significantly, suggesting
that the IRON galaxies have higher 20~cm radio powers because they are more
luminous sources at several wavelengths, not because their radio emission
is enhanced in some way.

\medskip
\begin{figure}[htb]
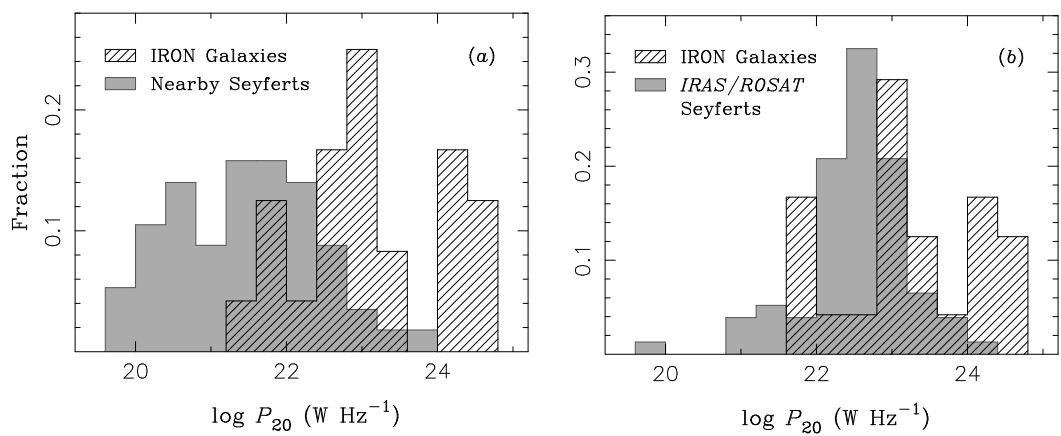

\epsfysize 2.2truein
\epsffile[69 370 532 745]{moran_fig1a.ps}
\vskip -2.20truein
\hskip 2.95truein
\epsfysize 2.2truein
\epsffile[99 370 531 745]{moran_fig1b.ps}
\caption{Rest-frame 20 cm radio luminosity distribution for the IRON
sample, compared to the radio luminosities of ($a$) nearby classical Seyfert
galaxies (Ulvestad \& Wilson 1989) and ($b$) classical Seyfert galaxies in
the {\sl IRAS-ROSAT\/} sample (Moran et al.\ 1996; Condon et al.\ 1998a).}
\end{figure}

{\it Radio Source Sizes.}  Most of the IRON galaxies are unresolved at
$\sim 1''$ resolution, confirming the conclusions of UAG that the nuclear
radio sources in NLS1s are compact.  However, in our 3.6 cm observations
(0\farcs25 resolution), two sources (IRAS 06269$-$0543 and Ark 564)
exhibit an unresolved core and what appears to be a small-scale ($\sim 1''$)
jet (Fig.~2).

\begin{figure}[htb]
\epsfysize 2.5truein
\epsffile[36 124 576 667]{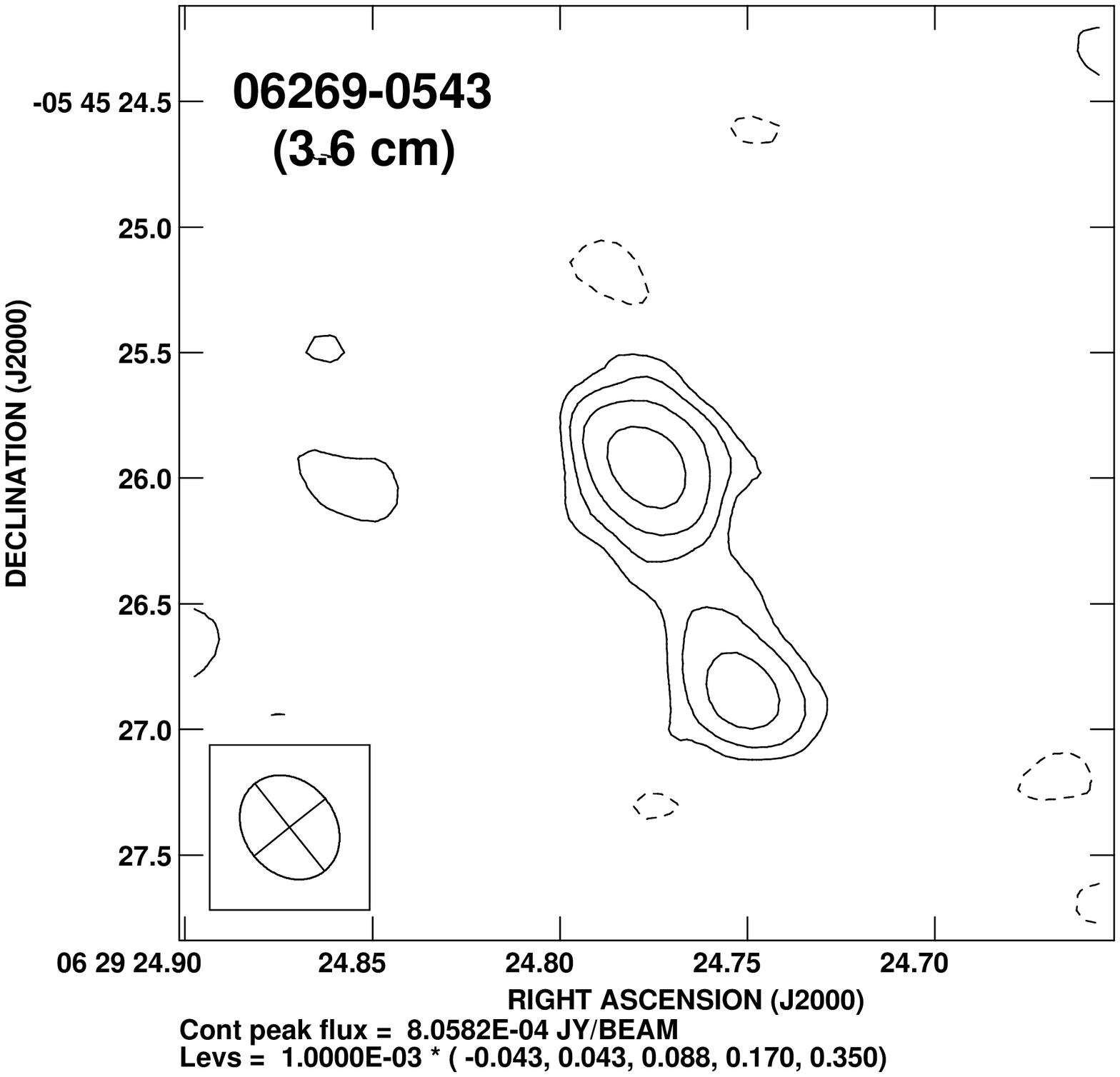}
\vskip -2.5truein
\hskip 2.90truein
\epsfysize 2.5truein
\epsffile[36 130 576 661]{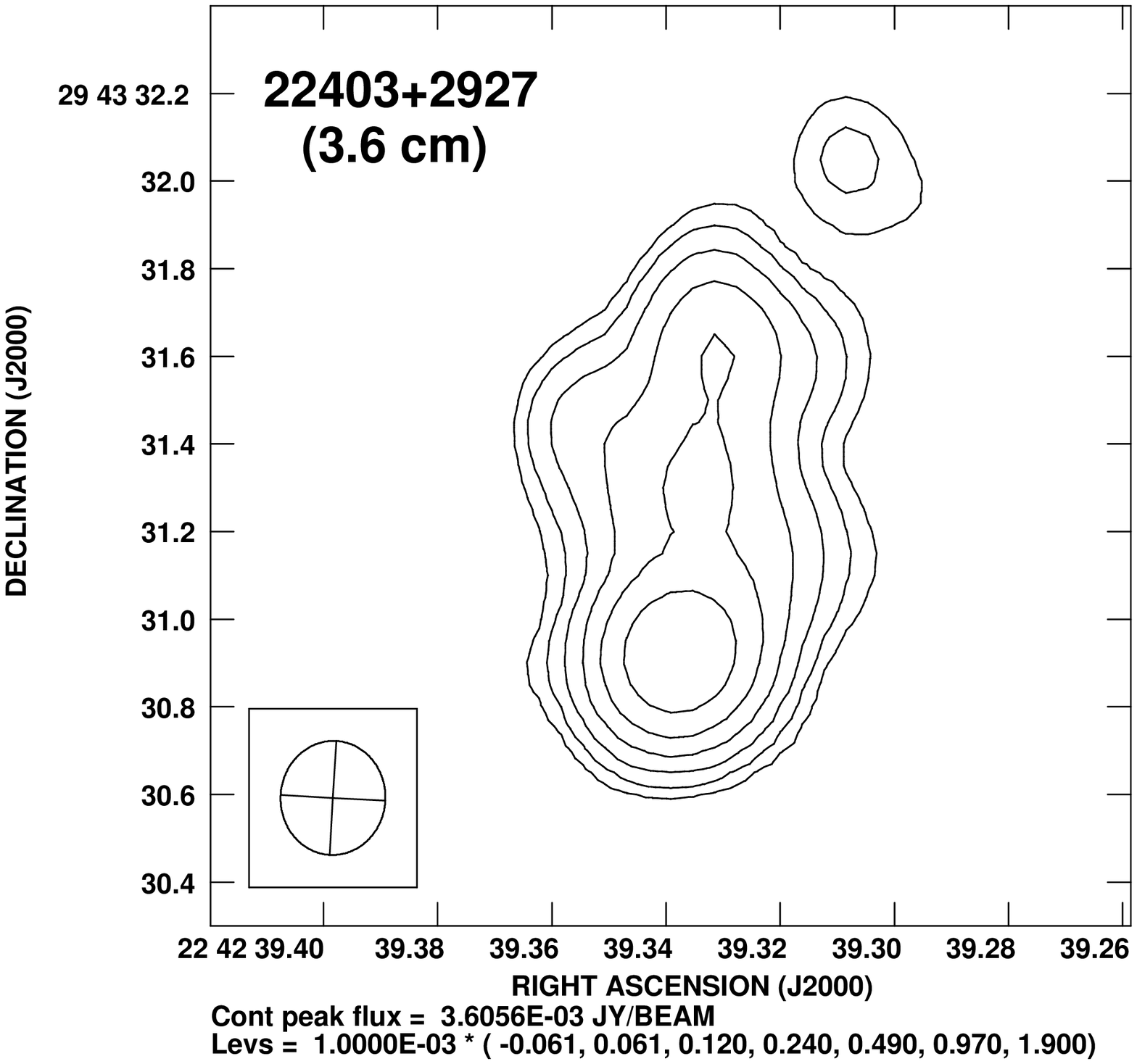}
\caption{VLA A-array images of two {\sl IRAS-ROSAT\/} NLS1s with evidence
for extended radio emission.  IRAS 22403+2927 is better known as Ark 564.}
\end{figure}

{\it Spectral Index Distribution.}  In Figure~3 we have plotted the radio
spectral index distribution for the IRON galaxies; also shown is the
distribution of spectral slopes for 59 of the classical Seyfert galaxies
in the distance-limited sample of Morganti et al.\ (1999).  Clearly, the IRON
galaxies tend to have significantly steeper radio spectra than the classical
Seyferts.  Only one of the IRON objects has a spectrum flatter than $\alpha
= 0.4$, and the bulk of the objects have $\alpha \approx 1.1 - 1.2$, well
out on the tail of the Morganti et al.\ Seyfert distribution.  One remarkable
source, IRAS 06269$-$0543, has a spectral index of $\alpha = 2.21$, which is
steeper by far than the spectrum of any core-dominated Seyfert galaxy or
radio-quiet quasar we are aware of.

{\it Variability.} It is difficult to evaluate the radio variability of the
IRON sample because of resolution effects associated with the different VLA
configurations used for the 20~cm observations.  However, a few galaxies
exhibit flux density differences that are not instrumental in nature,
including IRAS 06269$-$0543 (38\% variability) and IRAS 20181$-$2244
(18\% variability).

\section{Implications for the Physical Nature of NLS1s}
The nuclear radio sources in the IRON galaxies tend to be compact,
steep-spectrum, and, in a few cases, variable---three characteristics that
are rarely found together.  This unusual combination of properties can be
accounted for if most of the radio flux arises from a tiny ($\sim$~1 pc
diameter) region near the central engine of the active nucleus.  In this
scenario, the variability is not intrinsic to the source, but is caused by
``scintillation'' as the emission passes through the interstellar medium
of the Milky Way (e.g., Rickett 1990).  Due to the proximity of the
radio-emitting plasma to the intense optical/UV continuum source, the
cooling of the electrons is dominated by inverse-Compton scattering rather
than synchrotron emission, which steepens the radio spectrum.  In the case
of IRAS 06269$-$0543, our calculations indicate that the electron cooling
time would be very short in this picture, suggesting that relativistic
electrons are being continuously resupplied.  This might occur if the mass
accretion rate in this object is very high relative to the Eddington limit,
which has been suggested to explain the steep soft X-ray spectra and rapid,
large-amplitude X-ray variability observed in some NLS1s (Pounds et al.\
1995; Boller et al.\ 1996).  A thorough description of this hypothesis,
which can be tested with additional radio observations, is provided in
Moran et al.\ (2000).

\medskip
\begin{figure}[htb]
\epsfysize 2.4truein
\centerline{\epsffile[38 139 571 586]{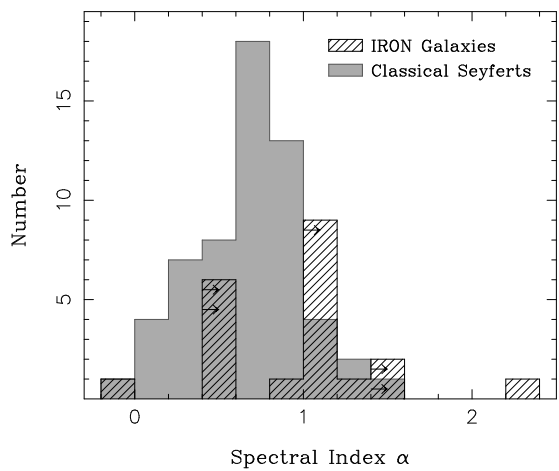}}
\caption{Radio spectral index distributions (assuming $S_{\nu}
\propto \nu^{-\alpha}$) for the IRON galaxies and 59 of the classical
type~1 and type~2 Seyfert galaxies in the distance-limited sample of
Morganti et al.\ (1999).}
\end{figure}



\end{document}